\documentclass[aps,prb,preprint,groupedaddress]{revtex4-1}
\usepackage{graphicx}
\usepackage{amsmath}
\usepackage{relsize}
\usepackage{hyperref}
\usepackage{multirow}
\usepackage{color}
\usepackage{hhline}
\usepackage{float}
\restylefloat{table}

\begin{document}

\title{Field-dependent specific heat of the canonical underdoped cuprate superconductor YBa$_2$Cu$_4$O$_8$}

\author{Jeffery L. Tallon$^{1,\dag}$, and John W. Loram$^{2,\ddag}$}

\affiliation{$^1$Robinson Research Institute, and MacDiarmid Institute for Advanced Materials and Nanotechnology, Victoria University of Wellington,
P.O. Box 33436, Lower Hutt 5046, New Zealand.}

\affiliation{$^2$Cavendish Laboratory, Cambridge University, Cambridge CB3 0HE, United Kingdom.}

\date{\today}
%


\maketitle

{\bf The cuprate superconductor YBa$_2$Cu$_4$O$_8$, in comparison with most other cuprates, has a stable stoichiometry, is largely free of defects and may be regarded as the canonical underdoped cuprate, displaying marked pseudogap behaviour and an associated distinct weakening of superconducting properties. This cuprate `pseudogap' manifests as a partial gap in the electronic density of states at the Fermi level and is observed in most spectroscopic properties. After several decades of intensive study it is widely believed that the pseudogap closes, mean-field like, near a characteristic temperature, $T^*$, which rises with decreasing hole concentration, $p$. Here, we report extensive field-dependent electronic specific heat studies on YBa$_2$Cu$_4$O$_8$ up to an unprecedented 400 K and show unequivocally that the pseudogap never closes, remaining open to at least 400 K where $T^*$ is typically presumed to be about 150 K. We show from the NMR Knight shift and the electronic entropy that the Wilson ratio is numerically consistent with a weakly-interacting Fermion system for the near-nodal states. And, from the field-dependent specific heat, we characterise the impact of fluctuations and impurity scattering on the thermodynamic properties.
}

Both the pseudogap \cite{Norman1,Timusk,Tstar} and the origins of superconductivity in the cuprates remain enigmatic and a source of continuing dispute, especially the former \cite{entrant}. Still there is no consensus as to pseudogap's phenomenology, at what doping the ground-state pseudogap ultimately vanishes, whether it really does close at $T^*$, whether this closure might be a thermodynamic phase transition \cite{Bourges,Xia,Hashimoto,He1,Shekhter,Sato} and whether it is causatively related to superconductivity \cite{entrant}. The electronic specific heat captures the entire spectrum of low-energy excitations and in principle can adjudicate in all these matters. The key experimental challenge is to separate the electronic term from the much larger phonon term. In many previous experiments \cite{Loram2,Loram1,Loram4} and in the present report this can be done using a differential technique in which the specific heat is measured relative to a reference sample which, if closely related to the sample itself, automatically backs off most of the  phonon contribution. The residual phonon contribution can be identified and removed by measuring a series of doping states in which the residual is found to scale linearly with the mass change of the doping process, usually changing oxygen content. Further details are given under {\bf Methods}.

\subsection*{Electronic specific heat coefficient}

\begin{figure}
\centering
\includegraphics[width=70mm]{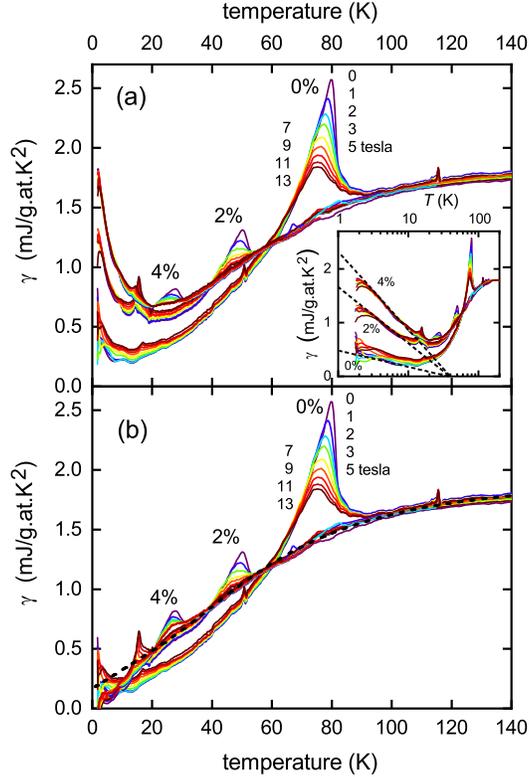}
\caption{\small
(a) The measured electronic specific heat coefficient $\gamma(T)$ for YBa$_2$Cu$_4$O$_8$ for 0\%, 2\% and 4\% planar Zn concentration at 0, 1, 2, 3, 5, 7, 9, 11 and 13 tesla. The same colour coding is used for each sample. Inset: the same data plotted versus $\ln(T)$ highlighting the impurity scattering term (dotted lines). (b) the same data with the impurity term subtracted. The black dashed curve is the fitted normal-state specific heat coefficient, $\gamma_{\textrm{n}}$ which satisfies entropy balance.  }
\label{gammaraw}
\end{figure}

The measured electronic specific heat coefficient, $\gamma(T) \equiv C_P(T)/T$ is shown in Fig.~\ref{gammaraw}(a) for YBa$_2$Cu$_4$O$_8$ (Y124) with 0\%, 2\% and 4\% planar Zn concentration at nine different applied fields, as annotated. As discussed later, the data extends to an unprecedented 400 K but we focus first on the transitions into the superconducting state. Because Zn substitutes only on the CuO$_2$ planes and not on the chains \cite{Williams1} these compositions correspond to YBa$_2$Cu$_{4-y}$Zn$_y$O$_8$ with $y=$ 0, 0.04 and 0.08. It is immediately evident that $T_c$ is rapidly suppressed by Zn substitution (as observed previously in other cuprates \cite{Tallon_scattering}) and along with this a rapid reduction in the jump height at $T_c$, $\Delta\gamma_{\textrm{c}}$. At the same time there is a rapid suppression in $\Delta\gamma_{\textrm{c}}$ with applied field that becomes more extreme in the Zn-substituted samples. Further, at low $T$, $\gamma(T)$ fans out to higher values with applied field in the `pure' sample but not appreciably in the doped samples. This is due to the Volovik effect \cite{Volovik} – the field-induced pairbreaking at the nodes due to Doppler shift of quasiparticle energies, as discussed below.

The most notable feature, however, is the low-$T$ upturn due to impurity scattering. The inset to Fig.~\ref{gammaraw}(a) shows $\gamma(T)$ plotted versus $\ln(T)$ and this reveals a common underlying energy scale given by the convergence of the dashed lines at 38 K. The dashed lines are subtracted from the raw data to give the $\gamma(T)$ versus $T$ plot in panel (b) and it is this that we proceed to analyze.  (There is a small anomaly at 18 K, present in the Zn-doped samples but very weak in the pure; and another at 120 K, present only in the pure sample. The sample variability indicates unidentified impurities and these anomalies are ignored in the following).

Our first task is to identify the normal-state coefficient, $\gamma_{\textrm{n}}$, that would occur in the absence of superconductivity. This is very much constrained by the displayed data for $\gamma(T)$ because $\gamma_{\textrm{n}}$ must follow each of the three data sets above their respective $T_{\textrm{c}}$ values. This is the black dashed curve. It is further tightly constrained by the requirement for entropy balance. Because the area under a $\gamma(T)$ curve is entropy then integrating $\gamma(T)$ from $T=0$ to some $T_0 > T_{\textrm{c}}$ must give the same result as integrating $\gamma_{\textrm{n}}(T)$ from $T=0$ to $T_0$. The fit function which satisfies these two requirements is


\begin{equation}
\gamma_{\textrm{n}}(T) = 1.93 \left[1 - 0.913 \tanh^{\alpha}\left(\frac{E^*}{2k_BT}\right) \frac{\ln[\cosh(E^*/2k_BT)]}{E^*/2k_BT}\right]
\label{gamma_n}
\end{equation}

\noindent where $E^* =13.44$ meV (or $T^* = E^*/k_B = 156$ K) and the exponent $\alpha = 1.7$. The general form of this equation for $\gamma(T)$ arises analytically from inserting into Eq.~\ref{spineq}, below, a triangular gap in the density of states (DOS) with a finite DOS at the Fermi level \cite{Tstar}. The amplitude 0.913 (being less than unity) reflects the finite DOS at $E_{\textrm{F}}$. This residual DOS is manifested in the finite value of $\gamma_{\textrm{n}}(0)=0.183$ mJ/g.at.K$^2$ and is a signature of the ungapped Fermi arcs, or hole pockets, of a reconstructed Fermi surface \cite{Storey2,Kunisada}.

$T^* = 156$ K is typical of values reported for Y124 from transport \cite{Bucher} and NMR relaxation \cite{Raffa} measurements but we emphasize this reflects an energy scale not a temperature \cite{entrant}. An important implication of Fig.~\ref{gammaraw} is that there is no coupling between superconductivity and the pseudogap in the sense that the onset of superconductivity does not weaken the pseudogap. This is evident from the fact that a single $\gamma_{\textrm{n}}(T)$ curve fits all three samples i.e. $\gamma_{\textrm{n}}(T)$ is the same for 4\% and 0\% Zn even in the temperature range below $T_{\textrm{c}}$ for 0\% Zn so that the onset of superconductivity in the latter case does not alter the underlying pseudogap energy scale, $E^*$. Close scrutiny of the $k$-dependent gap in Bi2212, as measured by angle-resolved photoelectron spectroscopy (ARPES) \cite{Vishik}, (which allows separation of the antinodal pseudogap from the nodal superconducting gap on the Fermi arcs) confirms that the pseudogap amplitude does not alter on cooling below $T_{\textrm{c}}$.

\subsection*{NMR Knight shift and entropy}

Next, we note that the spin susceptibility and electronic entropy are closely related. To see this consider the entropy for a weakly-interacting Fermi liquid \cite{Padamsee}:
\begin{equation}
S_{\textrm{n}} = -2k_B \int_{-\infty}^\infty \! [f\ln(f) + (1-f)\ln(1-f)] \, N(E) \,\mathrm{d}E
\label{entropyeq}
\end{equation}
\noindent where $f(E)$ is the Fermi function and $N(E)$ is the electronic DOS for one spin direction. This is just a weighted integral of the DOS with the `Fermi window' $[f\ln(f) + (1-f)\ln(1-f)]$.

On the other hand, the spin susceptibility for a weakly-interacting Fermion system is:
\begin{equation}
\chi_s = -2\mu_B^2 \int_{-\infty}^\infty \! \frac{\partial f(E)}{\partial E} \, N(E) \,\mathrm{d}E ,
\label{spineq}
\end{equation}
\noindent Therefore, like the entropy, the susceptibility is an integral of the DOS where the Fermi window is now the function $\partial f/\partial E$. It turns out that $T\partial f/\partial E$ is essentially identical to $[f\ln(f) + (1-f)\ln(1-f)]$ if $\chi_s$ in the former is stretched in temperature by a factor 1.187 \cite{Loram_IRC}. It is therefore not surprising that $S/T$ and $\chi_{\textrm{s}}$ are related. This relationship is expressed by the Wilson ratio, $a_{\textrm{W}}$, such that $S(T)/T = a_{\textrm{W}} \chi_{\textrm{s}}(T)$, where
\begin{equation}
a_W = \frac{1}{3\mu_0} \left(\frac{\pi k_B^2}{\mu_B}\right)^2
\label{Wilson}
\end{equation}

\begin{figure}
\centering
\includegraphics[width=70mm]{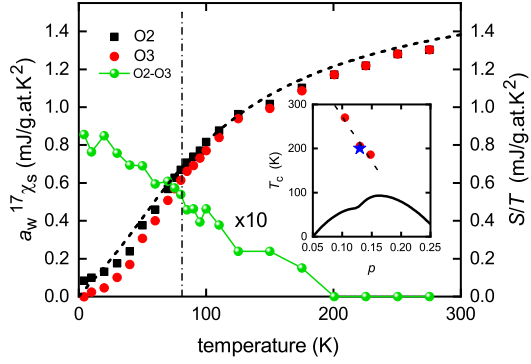}
\caption{\small
Dashed curve: The normal-state entropy coefficient, $S_{\textrm{n}}(T)/T$, obtained by integrating the dashed curve for $\gamma_{\textrm{n}}(T)$ in Fig.~\ref{gammaraw}(b). Red and black data points: the spin susceptibility, $\chi_{\textrm{s}}(T)$, calculated from the planar oxygen Knight shifts $^{17}K_{2,c}$ and $^{17}K_{3,c}$ associated with the O2 and O3 oxygen sites. $\chi_{\textrm{s}}(T)$ is expressed in entropy units by multiplying by the Wilson ratio, $a_W$, for weakly interacting Fermions. The green connected data points are the difference in O2 and O3 spin susceptibilities obtained from $^{17}K_{2,c} - ^{17}K_{3,c}$, showing the abrupt onset of nematic splitting at 200 K {\it within} the pseudogap state. The inset shows the cuprate phase diagram with the three red data points of Sato {\it et al.} \cite{Sato} marking the onset of nematicity. The blue star marks the onset of nematic splitting of the O2 and O3 Knight shifts.
}
\label{O_NMR}
\end{figure}

We will now test this relationship in the present case of Y124. By integrating $\gamma_{\textrm{n}}(T)$ from $T=0$ to $T$ we obtain the normal-state entropy and this is plotted as $S_{\textrm{n}}(T)/T$ by the black dashed curve in Fig.~\ref{O_NMR}. For fully-oxygenated Y123 $S_{\textrm{n}}(T)/T$ is essentially independent of temperature \cite{Loram3}, reflecting the fact that the pseudogap has closed at maximal doping ($p \approx 0.19$ holes/Cu). But for Y124 there is a large pseudogap present which suppresses $S/T$ at low $T$. This is also seen in the $T$-dependent NMR Knight shift which is linearly related to the spin susceptibility. To illustrate, we show in Fig.~\ref{O_NMR} the $^{17}$O Knight shift, referenced to the chemical shift, as reported by Tomeno {\it et al.} \cite{Tomeno}. These authors also report the bulk susceptibility as a function of the Knight shift, thus enabling calibration of the spin susceptibility from the Knight shift. As a final step we multiply the spin susceptibility by the Wilson ratio, $a_{\textrm{W}}$, in order to express the $T$-dependent part of the Knight shift in entropy/$T$ units.

It can be seen in Fig.~\ref{O_NMR} that, not only the shape, but the absolute magnitude concurs remarkably well with the derived $S_{\textrm{n}}(T)/T$ suggesting, as already noted for the bulk susceptibility \cite{Loram}, that the near-nodal states are consistent with a weakly interacting Fermionic system. Of especial interest is the fact that the O2 and O3 Knight shifts begin to diverge below 200 K. The difference in shift, $^{17}K_{2,c} - ^{17}K_{3,c}$, expressed in entropy units, is also shown in the figure ($\times$10). This shows an abrupt onset in nematicity, consistent with that reported by the Matsuda group \cite{Sato} using torque magnetometry. Its location at ($p$=0.13, $T$=200) is precisely consistent with the three data points in the ($p$,$T$) plane reported by the Matsuda group for Y123. (For the doping state of 0.13 see {\bf Materials}). These are plotted in the inset to Fig.~\ref{O_NMR} by the red data points. Their susceptibility data was presented as evidence for a thermodynamic ``phase transition at the onset of the pseudogap" however it is clear from Fig.~\ref{O_NMR} that the pseudogap is already open far above $T_{\textrm{nematic}}$, having already depleted half of the spin susceptibility. We observe no anomaly in $\gamma(T)$ at or near 200 K to suggest a phase transition. It must be very weak. We will see below that the pseudogap in fact extends at least to well above 400 K. Consequently this nematic phase transition occurs {\it within} a preexisting pseudogap state that extends far above and is not a transition {\it into} the pseudogap state, contrary to what has been claimed \cite{Sato}.

\begin{figure}
\centering
\includegraphics[width=70mm]{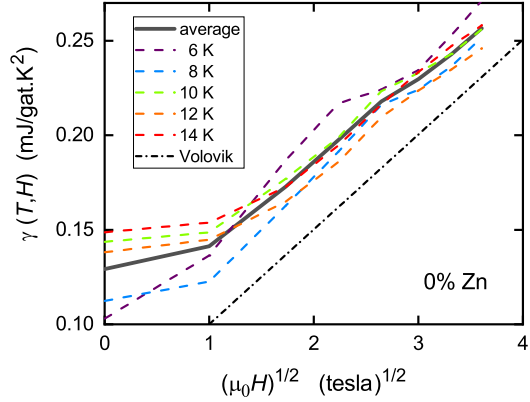}
\caption{\small
The measured low-temperature specific heat coefficient, $\gamma(T,H)$ for pure Y124 plotted as a function of field for $T$ = 6, 8, 10, 12 and 14 K. Above 1 tesla the data is essentially linear, consistent with the Volovik effect \cite{Volovik}. The black dash/dot line shows the expected slope for $\Delta_0 = 30$ meV.
}
\label{Volovik}
\end{figure}

\subsection*{Volovik effect}
We now consider the field-dependent low-$T$ behaviour of $\gamma(T)$ for the pure sample. Fig.~\ref{Volovik} shows $\gamma(T,H)$ plotted as a function of $\sqrt{\mu_0H}$ for $T$ = 6, 8, 10, 12 and 14 K. Above 1 tesla the behaviour is linear, consistent with the expected phenomenology of the Volovik effect. Such behaviour has been seen in the specific heats of single-crystal Y123 \cite{Junod} and La$_{2-x}$Sr$_x$CuO$_4$ \cite{Wang}, and in interlayer tunneling in Bi2212 \cite{Benseman}. For $d$-wave symmetry, in the superconducting state the finite ground-state specific heat coefficient is \cite{Wang}

\begin{equation}
\Delta\gamma(H,0) =  \frac{4k_{\textrm{B}}^2}{3\hbar}\sqrt{\frac{\pi}{\phi_0}} \frac{nV_{\textrm{M}}}{15d} \frac{a\hbar k_{\textrm{F}}}{2\Delta_0} \sqrt{\mu_0H} ,
\label{Volovik_eqn}
\end{equation} in units of mJ/g.at.K$^2$, where $V_{\textrm{M}}$ is the molar volume, $d$ the unit cell length, $\phi_0$ the flux quantum, $\Delta_0$ is the antinodal amplitude of the $d$-wave gap, and $k_{\textrm{F}}$ the Fermi wave vector at the node along the ($\pi$,$\pi$) direction. For $\Delta_0 =$ 30 meV the dash/dot line in Fig.~\ref{Volovik} shows the expected slope. The data is very consistent with this expectation. The averaged slope of 0.0443 mJ/g.at.K$^2$T$^{1/2}$ implies a gap amplitude of $\Delta_0 =$ 34 meV. We have used the value $k_{\textrm{F}} = 0.436\times 10^{10}$ m$^{-1} \, $ \cite{Sebastian}.

\begin{figure}
\centering
\includegraphics[width=70mm]{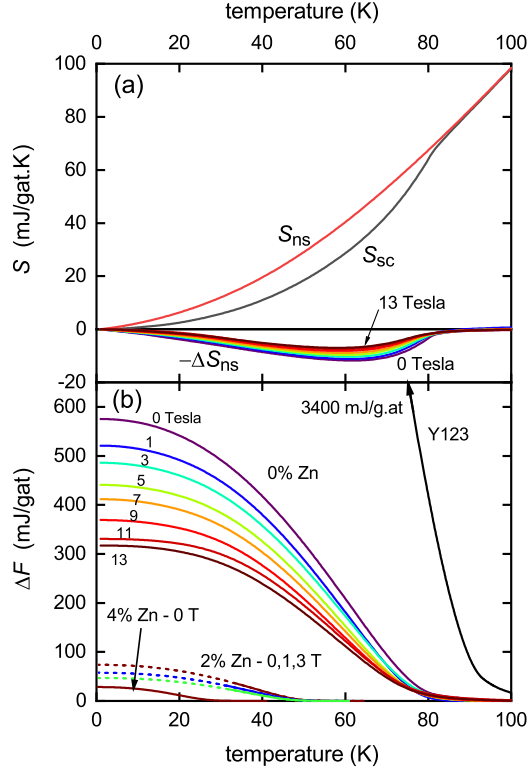}
\caption{\small
(a) The normal-state entropy, $S_{\textrm{n}}(T)$, obtained by integrating the black dashed curve in Fig.~\ref{gammaraw}(b) and the measured entropy, $S_{\textrm{s}}(T)$, for pure Y124 in zero field. The condensation entropy, $\Delta S_{\textrm{ns}} = S_{\textrm{n}}(T) - S_{\textrm{s}}(T)$ is shown underneath for $\mu_0 H$ = 0, 1, 3, 5, 7, 9, 11 and 13 T colour-coded as in (b). (b) The $T$-dependence of the condensation free energy $\Delta F_{\textrm{ns}} = F_{\textrm{n}} - F_{\textrm{s}}$ for pure Y124 obtained by integrating $\Delta S_{\textrm{ns}}(T)$ from above $T_{\textrm{c}}$ for each field as annotated. Below these are shown the condensation energy for the 2\% Zn and 4\% Zn samples showing a rapid suppression with scattering and applied field. The condensation energy for fully-oxygenated Y123 is also shown and is much higher.
}
\label{condensation}
\end{figure}

\subsection*{Free energy and superconducting gap}

Fig.~\ref{condensation}(a) shows the normal-state and superconducting state entropy below 100 K obtained by integrating $\gamma_{\textrm{n}}(T)$ and $\gamma_{\textrm{s}}(T)$, respectively (as displayed in Fig.~\ref{gammaraw}(b)) from 0 to $T$. The curves shown are for pure Y124 in zero field and they are denoted $S_{\textrm{n}}(T)$ and $S_{\textrm{s}}(T)$, respectively. The difference is the condensation entropy $\Delta S_{\textrm{ns}} = S_{\textrm{n}}(T) - S_{\textrm{s}}(T)$. This is plotted underneath for fields of $\mu_0 H$ = 0, 1, 3, 5, 7, 9, 11 and 13 T, colour-coded as in panel (b). Clearly the condensation entropy is rapidly suppressed in field.  Another feature of note is the presence of fluctuations around $T_{\textrm{c}}$ which broadens the transition somewhat. This will be discussed later.

\begin{figure}
\centering
\includegraphics[width=70mm]{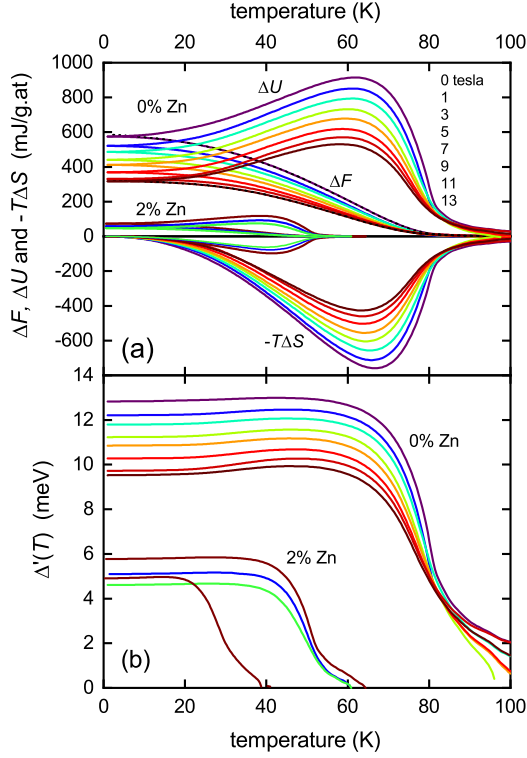}
\caption{\small
(a) The $T$-dependence of $\Delta F_{\textrm{ns}}(T)$ and its components $\Delta U_{\textrm{ns}}(T)$ and $-T\Delta S_{\textrm{ns}}(T)$ calculated from Eq.~\ref{free-energy} using $\gamma_{\textrm{n}} - \gamma_{\textrm{s}}$ as given in Fig.~\ref{gammaraw}. Most notable is the persistence of fluctuations high above $T_{\textrm{c}}$ in both $\Delta U_{\textrm{ns}}(T)$ and $-T\Delta S_{\textrm{ns}}(T)$ and the almost complete suppression in $\Delta F_{\textrm{ns}}(T)$.
(b) The superconducting order parameter, $\Delta^{\prime}(T)$, calculated from $2\Delta U_{\textrm{ns}}(T) - T\Delta S_{\textrm{ns}}(T)$ using Eq.~\ref{energygap} for 0\% and 2\% Zn at fields given by the colour coding in panel (a).
}
\label{DFDU}
\end{figure}

By integrating $\Delta S_{\textrm{ns}}(T)$ from a temperature $T_0$, sufficiently above the fluctuation regime that $\Delta S_{\textrm{ns}}(T_0) = 0$, down to a temperature, $T$, one obtains the condensation free energy $\Delta F_{\textrm{ns}}(T)$. However, a more efficient way of calculating the condensation free energy using just a single integration is given by:

\begin{equation}
-\Delta F_{\textrm{ns}}(T) =  \int_{T_0}^T \! T \, \Delta\gamma_{\textrm{ns}}(T) \,\mathrm{d}T - T\,\int_{T_0}^T \!  \Delta\gamma_{\textrm{ns}}(T) \,\mathrm{d}T ,
\label{free-energy}
\end{equation}
\noindent where the first term is the condensation internal energy, $-\Delta U_{\textrm{ns}}(T)$, and the second term is the condensation entropy term, $T\Delta S_{\textrm{ns}}(T)$. The condensation free energy calculated in this way is plotted in Fig.~\ref{condensation}(b) for 0\%, 2\% and 4\% Zn and for the various annotated fields. $\Delta F_{\textrm{ns}}(T)$ and its components $\Delta U_{\textrm{ns}}(T)$ and $-T\Delta S_{\textrm{ns}}(T)$ are plotted in Fig.~\ref{DFDU}.  Also plotted in Fig.~\ref{condensation}(b) is the condensation energy for fully oxygenated Y123 which rises to a ground-state value of 3400 mJ/g.at - a full six-fold greater than for pure Y124. This shows the full impact of the pseudogap for Y124 in weakening superconductivity. Also evident is the dramatic effect of impurity scattering in further reducing the condensation energy (which is particularly marked in underdoped cuprates where the pseudogap is present \cite{Tallon_scattering}). $\Delta F_{\textrm{ns}}(0)$ for the 4\% Zn-doped sample in zero field is just 28 mJ/g.at - 125 times smaller than for pure Y123. The curves for 2\% Zn are dashed below 30 K and this is because $\Delta S_{\textrm{ns}}(T)$ is a little noisy at low $T$ and in some cases does not fall exactly to zero, as it must. We find the first 20 K of the data scales precisely with $\Delta S_{\textrm{ns}}(T)$ for the 0\% Zn sample, and so we assumed that this scaling continues down to $T=0$ thus enforcing $\Delta S_{\textrm{ns}}(T)$ to fall to zero as $T \rightarrow 0$. Any errors introduced are very small - of the order of the thickness of the curves and of no consequence in the following analysis.

As mentioned, Fig.~\ref{DFDU}(a) shows $\Delta F_{\textrm{ns}}(T)$ and its components $\Delta U_{\textrm{ns}}(T)$ and $-T\Delta S_{\textrm{ns}}(T)$. It is striking that fluctuations persist high above $T_{\textrm{c}}$ in both $\Delta U_{\textrm{ns}}(T)$ and $-T\Delta S_{\textrm{ns}}(T)$ while they are almost completely cancelled in $\Delta F_{\textrm{ns}}(T)$. The superconducting gap function $\Delta(T)$ may be calculated from these components of the free energy using \cite{Tallon2}

\begin{align}
\zeta N(0) \Delta(T)^2 &= 2\Delta F_{\textrm{ns}}(T) + T\Delta S_{\textrm{ns}}(T) \nonumber \\
&\equiv 2\Delta U_{\textrm{ns}}(T) - T\Delta S_{\textrm{ns}}(T)
\label{energygap}
\end{align}
\noindent where $\zeta$ = 1 for $s$-wave and 1/2 for $d$-wave. The calculated $\Delta(T)$ values are plotted in Fig.~\ref{DFDU}(b). The gap is the magnitude of the order parameter, $\Delta^{\prime}$, rather than the spectral gap, $\Delta$, which is higher \cite{Loram1}. Roughly speaking, $\Delta^{\prime}_0 = \sqrt{\Delta_0^2 - E^{*2}}$ \cite{Loram1}. From the impurity suppression of superfluid density \cite{Tallon7} we estimate $\Delta_0 \approx 23.4$ meV while from the high-$T$ entropy suppression we determine $E^* \approx 19.1$ meV. The quadratic relation above then implies $\Delta_0^{\prime} = 13.5$ meV, very consistent with the values in Fig.~\ref{DFDU}(b). The values of $\Delta^{\prime}(T)$ are seen to descend towards zero at $T_{\textrm{c}}$ then persist with a more slow decline above $T_{\textrm{c}}$. Here the pairing is incoherent \cite{Storey1,Kondo} and is a feature of the strong superconducting fluctuations above $T_{\textrm{c}}$.

\subsection*{Comparison of Y124 with Y123}
\begin{figure}
\centering
\includegraphics[width=70mm]{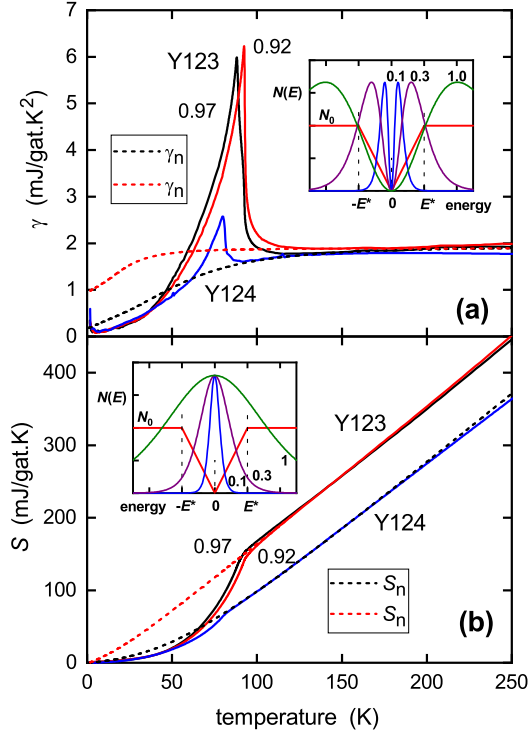}
\caption{\small
Comparison of (a) $\gamma(T)$ and (b) $S(T)$ for Y124 with nearly fully-oxygenated Y123. At highest oxygenation, $x=0.97$, a double transition (see side-hump) is observed while a small depletion, $x=0.92$, sees these merge into a single transition with a larger peak size. Above $T_{\textrm{c}}$ the $\gamma(T)$ curves merge while for $S(T)$ they remain separated. The inset in (a) shows the double-peaked Fermi window, $\partial \mathcal{F}/\partial T$, for calculating $\gamma(T)$ using Eq.~\ref{entropyeq}, where $\mathcal{F}(E,T) = [f\ln(f) + (1-f)\ln(1-f)]$. The window is centred on a normal-state triangular gap in the DOS (red lines) located at $E_{\textrm{F}}$. The blue, purple and olive-green curves show $\partial \mathcal{F}/\partial T$ for temperatures $k_{\textrm{B}}T = 0.1E^*$, $0.3E^*$ and $1.0E^*$, respectively. The inset in (b) shows the single-peaked Fermi window, $\mathcal{F}(E,T)$, for calculating entropy, again centred on a triangular gap in the DOS.
}
\label{compare}
\end{figure}

It is highly instructive to compare the {\it measured} data for Y124 with that for Y123. This is shown for $\gamma(T)$ in Fig.~\ref{compare}(a) and for $S(T)$ in Fig.~\ref{compare}(b). The Y123 data is taken from ref. \cite{Loram_IRC} which, in contrast to earlier reports \cite{Loram1,Cooper1}, includes a small correction for the background DOS in the undoped state. We also show the entropy-conserving normal-state functions as well. Two features are prominent. (i) the size of the specific heat jump for Y124 is much smaller than for Y123 due to the presence of the pseudogap in the former. (ii) while the $\gamma(T)$ curves converge above $T_{\textrm{c}}$ the $S(T)$ curves remain separated and parallel to the highest temperature. This is clear evidence that the pseudogap remains present in Y124 to the highest temperatures measured - here 400 K.

\begin{figure}
\centering
\includegraphics[width=70mm]{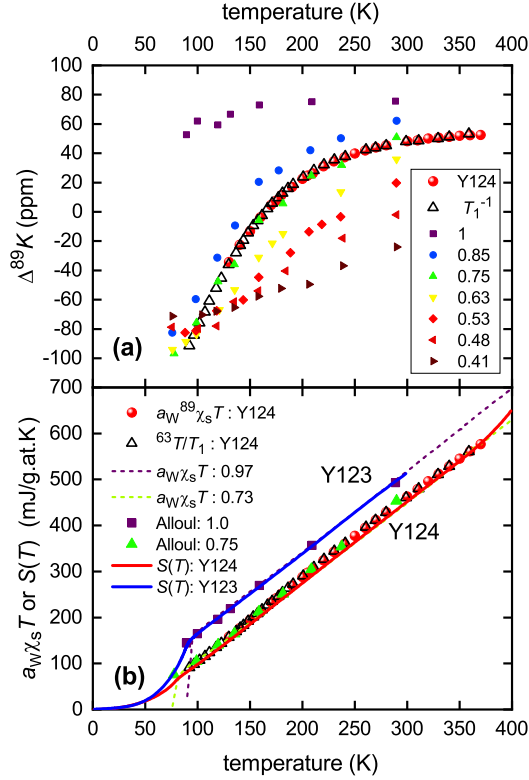}
\caption{\small
(a) $^{89}$Y NMR Knight shift for YBa$_2$Cu$_3$O$_{6+x}$ as reported by Alloul {\it et al.} \cite{Alloul} with $x$ values annotated. Also shown is the MAS $^{89}$Y NMR Knight shift and the scaled $T$ variation of $1/T_1$ for YBa$_2$Cu$_4$O$_8$ as reported by Williams {\it et  al.} \cite{Williams2}. (b) Y123: Entropy $S$ (blue solid curve \cite{Loram_IRC}), $a_{\textrm{W}}\,^{89}\chi_{\textrm{s}}T$ (purple squares \cite{Alloul}) and bulk susceptibility $a_{\textrm{W}}\chi_{\textrm{s}}T$ (purple dashed curve \cite{Loram_IRC}) for Y123 at nearly full oxygenation. Y124: Entropy $S$ (red solid curve - this work), $a_W\,^{89}\chi_{\textrm{s}}T$ (red spheres \cite{Williams}) and $T/^{63}T_1$ (open black triangles \cite{Raffa,Williams2}) for Y124. Then, as a proxy for Y124: $a_W\,^{89}\chi_{\textrm{s}}T$ for Y123 with $x=0.75$  (solid green triangles - Alloul \cite{Alloul}); and $a_W\chi_{\textrm{s}}T$ for Y123 with $x=0.73$ (dashed green curve - Loram \cite{Loram_IRC}).
}
\label{Y89_Ks}
\end{figure}

To see this, consider the Fermi window for the entropy given in Eq.~\ref{entropyeq}. This is the single-peaked function shown in the inset to Fig.~\ref{compare}(b). Three different temperatures of 0.1, 0.3 and 1.0 $E^*/k_{\textrm{B}}$ are displayed, where $E^*$ is the magnitude of the triangular gap shown in the figure as a representation of the pseudogap. Provided that the pseudogap remains open, this Fermi window always sees the gap and at higher temperatures loses a fixed fraction of states so that $S(T)$ is displaced down in parallel fashion relative to that for Y123 - as evidenced in Fig.~\ref{compare}(b), and again in Fig.~\ref{Y89_Ks}(b). In contrast, because $\gamma \equiv \partial S/\partial T$, the Fermi window for $\gamma(T)$ is the $T$-derivative of that for $S(T)$. This is a double-peaked function as shown in the inset to Fig~\ref{compare}(a). It can be seen that at high enough temperature this Fermi window falls outside of the gap. Therefore $\gamma(T)$ recovers its full ungapped magnitude at high $T$, despite the presence of the gap. The data in Fig.~\ref{compare}(a) and (b) are entirely consistent with this picture. The two systems have essentially the same background DOS. In the normal-state Y123 is ungapped while Y124 is gapped to the highest temperature as evidenced by the parallel suppression of $S(T)$. If the gap were to close with increasing $T$ then $S(T)$ for Y124 would recover to that for Y123. It does not. By extrapolating $S(T)$ back to the ordinate axis one can read off the normal-state gap magnitude. For a triangular gap, as in the insets to Fig.~\ref{compare}, but with a finite DOS of $N_1$ at $E_{\textrm{F}}$ and a constant DOS of $N_0$ above $E^*$ this negative intercept is $2\ln2 \,k_{\textrm{B}}(N_0 - N_1)V_{\textrm{M}} E^*$, where $V_{\textrm{M}}$ is the molar volume. The finite value of $\gamma_{\textrm{n}}$ at $T=0$ is given by $\gamma_{\textrm{n}}^0 = 4 \ln2 \times2.374 k_{\textrm{B}}^2 N_1 V_{\textrm{M}}$, while at high $T$ we have $\gamma_{\textrm{n}} =(2/3)\pi^2 k_{\textrm{B}}^2 N_0 V_{\textrm{M}}$. From these we obtain $E^* = 19.1$ meV and $N_0 = 6.14\times10^{-3}$ states/meV/cell.

\subsection*{High-temperature susceptibility and entropy}
We now extend this comparison to 400 K, a range never previously achieved in differential measurements. We combine this data with $^{89}$Y Knight shift data, bulk susceptibility measurements and $1/^{63}T_1$ data in such a way that demonstrates, individually and collectively, the persistence of the pseudogap to 400 K and beyond. $^{89}$K$_{\textrm{s}}(T)$ probes the spin susceptibility of the CuO$_2$ planes that sandwich the Y atom in Y123 and Y124 \cite{Alloul}. The $^{89}$Y nucleus has the additional benefit of having no quadrupole moment so there is no quadrupole splitting of the resonance arising from electric field gradients. Further, the use of magic-angle spinning (MAS) enables extremely narrow line widths as will be used below \cite{Williams}. The $1/^{63}T_1$ relaxation rate is a weighted sum over $\textrm{\bf q}$ of the imaginary part of the spin susceptibility, $\chi^{\prime\prime}(\textrm{\bf q}, \,\omega)$, where, for the $^{63}$Cu nucleus, the weighting form factor is strongly enhanced near the antiferromagnetic wave vector, $\textrm{\bf q}=(\pi,\pi)$ \cite{Crossover} and hence $1/^{63}T_1$ is dominated by the antinodal pseudogap.

Fig.~\ref{Y89_Ks}(a) shows the $^{89}$Y Knight shift $^{89}$K$_{\textrm{s}}(T)$ for YBa$_2$Cu$_3$O$_{6+x}$ as reported by Alloul {\it et al.} \cite{Alloul}, with values of $x$ annotated. Also plotted is MAS $^{89}$K$_{\textrm{s}}$ for samples of Y124 from our laboratory \cite{Williams2} (red spheres), along with the $1/^{63}T_1$ data from Raffa {\it et al.} \cite{Raffa} (up triangles) which we showed \cite{Williams2} scales precisely with $^{89}$K$_{\textrm{s}}$, as can be seen. (The conversion scale is $^{89}$K$_{\textrm{s}} = 48.8\times 1/^{63}T_1 - 164.5$.) Notably, there is an excellent match over the entire temperature range with Alloul's data for $x=0.75$ (green up-triangles) where the doping state (0.13) and $T_{\textrm{c}}$ for Y123 are much the same as those of Y124. As in Fig.~\ref{O_NMR}, we convert $^{89}$K$_{\textrm{s}}$ to spin susceptibility using the calibration of Alloul {\it et al.} \cite{Alloul} (see {\bf Methods} for more detail) and multiply by $a_{\textrm{W}}$ to express in $S/T$ units.  We find an excellent agreement between Alloul's $^{89}$K$_{\textrm{s}}$ and the measured entropy for Y123 across a wide range of doping and temperature (see Fig. S4 in Supplementary Information, SI).

In Fig.~\ref{Y89_Ks}(b) we assemble four distinct data sets ($S$, $^{89}$K$_{\textrm{s}}$, $1/^{63}T_1$, and $\chi_{\textrm{s}}$ from the bulk susceptibility) for Y123 at near full oxygenation ($x=0.97$), and for Y124. All are expressed in entropy units, in this case using the factor $a_W T$ to convert susceptibilities (including the $1/^{63}T_1$ data expressed as a susceptibility). The $\chi_{\textrm{s}}$ data is shown for Y123 with $x=0.97$ and $x=0.73$ (purple and green dashed curves, respectively, the latter as a proxy for Y124) and is taken from Loram {\it et al.} \cite{Loram_IRC}. Evidently $S$ (blue solid curve), $a_W\,^{89}\chi_{\textrm{s}}T$ (purple squares) and $a_W\chi_{\textrm{s}}T$ (purple dashed curve) for Y123 at full oxygenation all track linearly to the origin indicating the absence of the pseudogap. For Y124 on the other hand, $S$ (red solid curve - this work), $a_W\,^{89}\chi_{\textrm{s}}T$ (red spheres) and $T/^{63}T_1$ (open black triangles), together with Alloul's $a_W\,^{89}\chi_{\textrm{s}}T$ (solid green triangles) for Y123 with $x=0.75$ and $a_W\chi_{\textrm{s}}T$ (dashed green curve) for Y123 with $x=0.73$, both as proxies for Y124, all reveal a linear high-temperature behaviour that extrapolates to a negative intercept on the $y$-axis indicating the presence of a gap - the pseudogap. For Y124 the entropy curve almost completely overlays the bulk susceptibility green-dashed curve which is barely visible, so the figure is reproduced in the SI with the green-dashed curve overlaying the entropy. Evidently, the agreement over the full temperature range is excellent. Importantly, our entropy data extends to 400 K as does the $\chi_{\textrm{s}}$ data, and the $^{89}$K$_{\textrm{s}}$ data extends to 370 K. Fig.~\ref{Y89_Ks}(b) represents the central result of this work. There is no indication, at any temperature, of the entropy recovering to the gap-less curve observed for fully-oxygenated Y123 that would signify the closing of the pseudogap at, or around, some $T^*$ value. The small upturn in $S(T)$ near 400 K simply represents the limitations of the present differential technique at such a high temperature and is not seen in the $\chi_{\textrm{s}}$ data. We conclude that the pseudogap does not close at some postulated $T^*$ in the range 150 to 200 K but remains open to the highest temperature investigated - 400 K. A similar conclusion has recently been drawn from $1/^{17}T_1$ planar oxygen NMR relaxation data for a number of cuprates \cite{Haase}. We showed the same long ago \cite{Tstar} for the in-plane resistivity and similarly for the $c$-axis resistivity \cite{entrant,Bernhard1}.

\begin{figure}
\centering
\includegraphics[width=70mm]{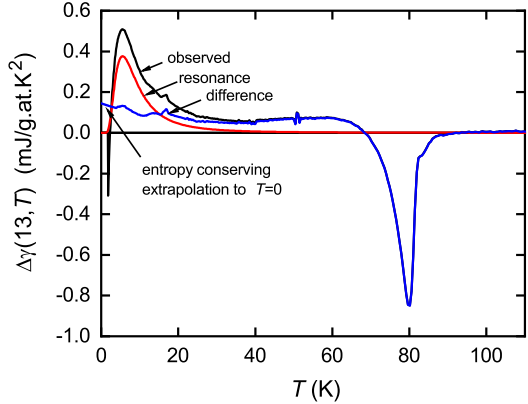}
\caption{\small
Black curve: The as-measured field-induced differential specific heat coefficient, $\Delta\gamma(13,T) = \gamma(13,T) - \gamma(0,T)$. Red curve: resonance $\gamma(T)$ as calculated using Eq.~\ref{resonance}. The blue curve is the difference, which, below 2.8 K, is extrapolated to a finite value as $T \rightarrow 0$ to ensure overall entropy conservation.
}
\label{resonance_g}
\end{figure}

\subsection*{Scattering resonance}
Finally, we wish to discuss the upturn in the raw $\gamma(T)$ data at low $T$ seen in Fig.~\ref{gammaraw}(a). This is in fact a peak rather than an upturn and may be identified with an impurity resonance \cite{Pan}. Scanning tunneling spectroscopy (STS) measurements in lightly Zn-doped Bi$_2$Sr$_2$CaCu$_2$O$_{8+\delta}$ reveal resonance spots in spatial maps at low energy and low temperature \cite{Pan}. Away from these spots, tunneling spectra reveal a well-formed $d$-wave superconducting gap with sharp coherence peaks. Tunneling spectra collected on the spots (the location of individual Zn atoms) show a nearly full suppression of both the gap and the coherence peaks with, instead, a sharp resonance appearing at $\varepsilon_{\textrm{r}} = - 1.5$ meV. To calculate the entropy contribution arising from this resonance we replace the DOS in Eq.~\ref{entropyeq} by a delta function, $N(E)=N_{\textrm{r}}\delta(\varepsilon-\varepsilon_{\textrm{r}})$. The equation integrates to give:
\begin{equation}
\Delta\gamma_{\textrm{res}} = A_{\textrm{res}} \left(\frac{\varepsilon_{\textrm{r}}}{2k_{\textrm{B}}T}\right)^3 \cosh^{-2} \left(\frac{\varepsilon_{\textrm{r}}}{2k_{\textrm{B}}T}\right)
\label{resonance}
\end{equation}
\noindent where the amplitude $A_{\textrm{res}} = \left(4k_{\textrm{B}}^2/\varepsilon_{\textrm{r}}\right) N_{\textrm{r}}$.

In view of the relationship between entropy and spin susceptibility discussed above, it is highly instructive to contrast this resonance component of $\gamma(T)$ with that of the susceptibility. Again, replacing the DOS in Eq.~\ref{spineq} by $N(E)=N_{\textrm{r}}\delta(\varepsilon-\varepsilon_{\textrm{r}})$ we find:
\begin{equation}
\chi_{\textrm{s,res}} = B_{\textrm{res}} \left(\frac{\varepsilon_{\textrm{r}}}{2k_{\textrm{B}}T}\right) \cosh^{-2} \left(\frac{\varepsilon_{\textrm{r}}}{2k_{\textrm{B}}T}\right)
\label{resonance_chi}
\end{equation}
\noindent where the amplitude $B_{\textrm{res}} = \left(\mu_{\textrm{B}}^2/2\varepsilon_{\textrm{r}}\right) N_{\textrm{r}}$. The interesting point in relation to Eqs.~\ref{resonance} and ~\ref{resonance_chi} is that $\Delta\gamma_{\textrm{res}}$ falls off rapidly as $T^{-3}$ while $\chi_{\textrm{s,res}}$ falls off more slowly as $T^{-1}$. This is borne out by our experimental data. Experimental evidence from the magnetic susceptibility for a Zn-induced resonance within the pseudogap was discussed previously in relation to Y124 \cite{Williams1} and La$_{2-x}$Sr$_x$CuO$_4$ \cite{Islam}.

Fig.~\ref{resonance_g} shows the as-measured field-induced change in specific heat coefficient, $\Delta\gamma(13,T) = \gamma(13,T) - \gamma(0,T)$. Recall that this difference contains no correction for the residual phonon contribution so is free of any imposed model. As well as the suppression of fluctuations around the specific heat anomaly near $T_{\textrm{c}}$ the low-temperature resonance is evident. We fit this using Eq.~\ref{resonance} with the parameters $A_{\textrm{res}} = 0.61$ mJ/g.at.K$^2$ and $\varepsilon_{\textrm{r}} = -1.55$ meV, the latter value being nicely consistent with the STS result for Bi$_2$Sr$_2$CaCu$_2$O$_{8+\delta}$ \cite{Pan}. This fit is the red curve in Fig.~\ref{resonance_g} and the difference is shown by the blue curve. The calculated resonance response is an excellent fit and shows the rapid decay at higher temperatures associated with the $T^{-3}$ tail. The difference is close to entropy conserving and requires a straightforward extrapolation below 2.8 K of its trend above 2.8 K to achieve exact entropy conservation. This rapid decay of the resonance in $\gamma(T)$ contrasts the predicted much slower $T^{-1}$ decay in the resonance part of the spin susceptibility. We have previously investigated the $^{89}$Y NMR Knight shift in Zn-doped Y124 \cite{Williams1}. The Zn resonance contribution to the spin susceptibility is seen in a satellite peak which has a slowly-decaying Curie temperature dependence observable all the way up to 300 K, thus nicely confirming the behaviour predicted by Eq.~\ref{resonance_chi}.

\subsection*{Conclusions}
In conclusion, we have measured the electronic specific heat of YBa$_2$Cu$_{4-x}$Zn$_{x}$O$_8$ using a precision differential technique that allows separation of the electronic term from the lattice term up to an unprecedented 400 K. The pure sample reveals the expected Volovik effect which is fully suppressed in the Zn-doped samples. We show that the pseudogap, characteristic of underdoped cuprates, always remains open to above 400 K, far above the nominal pseudogap onset temperature usually proposed, $T^* \approx$ 150 - 200 K. Weak thermodynamic transitions reported at $T^*$, for example the onset of susceptibility nematicity, occur {\it within} the already fully established pseudogap and are not transitions {\it into} the pseudogap state as widely claimed. The spin susceptibility, derived from the $^{17}$O and $^{89}$Y Knight shift and expressed in entropy units, is numerically the same as the electronic entropy divided by temperature, indicating that the near-nodal states are those of weakly-interacting Fermions. We derive the field-dependent condensation energy and superconducting energy gap. These expose the presence of strong pairing fluctuations extending well above $T_{\textrm{c}}$ as well as canonical impurity scattering behaviour, including the expected low-temperature resonance response which in the entropy channel decays rapidly as $T^{-3}$, but in the spin channel decays slowly as $T^{-1}$. The measurements and analysis reveal the remarkable utility of the differential specific heat technique in exposing the rich physics of strongly-correlated electronic materials.

\subsection{Methods}
{\bf Materials.} The samples were prepared from stoichiometric proportions of high purity Y$_2$O$_3$, dried Ba(NO$_3$)$_2$, ZnO and CuO, pressed as pellets and reacted for 16 hours at 935$^{\circ}$C under an oxygen pressure of 60 bar. The samples were ground finely and the process repeated three more times. X-ray diffraction revealed single-phase YBa$_2$Cu$_4$O$_8$ (see Fig. S1) with impurity less than 2\% (only CuO identified). Lattice parameters were found to be $a = 0.3842$ nm, $b = 0.3870$ nm, and $c = 2.7235$ nm under A{\it mmm} symmetry with a preferred alignment of the $c$-axis normal to the plane of the pellets. The $T_{\textrm{c}}$ values, determined from sharp diamagnetic onset were 81.2 K (0\% Zn), 51.8 K (2\% Zn) and 29.2 K (4\% Zn). The quoted Zn concentrations are those referred to the CuO$_2$ plane as essentially no Zn resides on the chains. Thus 2\% Zn refers to the composition YBa$_2$Cu$_{3.96}$Zn$_{0.04}$O$_8$ and 4\% refers to YBa$_2$Cu$_{3.92}$Zn$_{0.08}$O$_8$. Thermoelectric power measurements at 290 K give 6.95 $\mu$V/K (0\% Zn), 6.89 $\mu$V/K (2\% Zn) and 6.93 $\mu$V/K (4\% Zn). From the correlation of thermoelectric power with doping \cite{Obertelli} this amounts to essentially identical doping states of 0.130 holes/Cu for each.

The suppression of $T_{\textrm{c}}$ with Zn substitution is very much in line with that for underdoped Y123 at the same doping state. Fig. S2 shows $T_{\textrm{c}}$ as a function of doping for 0, 2, 4 and 6\% planar Zn substitution for Y$_{0.8}$Ca$_{0.2}$Ba$_2$Cu$_3$O$_{7-\delta}$ while the red stars show the data for Zn-substituted Y124. They are very consistent. The rapid suppression of $T_{\textrm{c}}$ for underdoped samples (open symbols) compared with overdoped (filled symbols) is a signature of the pseudogap which lowers the DOS at the Fermi level and hence raises the scattering rate \cite{Tallon_scattering}. Fig. S3 shows $T_{\textrm{c}}$ as a function of planar Zn concentration for Y$_{0.8}$Ca$_{0.2}$Ba$_2$Cu$_3$O$_{7-\delta}$ at various doping states while red stars show the same data for Y124. Again they are very consistent. Typically, for higher Zn concentrations, the $T_{\textrm{c}}$ value sits higher than that expected from Abrikosov-Gorkov pairbreaking due to the statistical overlap of nearest neighbours to the Zn substituent \cite{Tallon_scattering}.

{\bf Spin susceptibility and Knight shift.} The spin susceptibility, $\chi_{\textrm{s}}$, is related to the bulk magnetic susceptibility, $\chi_{\textrm{m}}$, by $\chi_{\textrm{m}} = \chi_{\textrm{s}} + \chi_0$ where the constant $\chi_0$ comprises diamagnetic and van Vleck terms and is evaluated for Y123 by Alloul {\it et al.} \cite{Alloul}. The measured Knight shift, $^{89}$K$_{\textrm{s}}$, is given by $^{89}$K$_{\textrm{s}}$ = $\sigma_0(x) + a(x)\,^{89}\chi_{\textrm{s}}$, where $\sigma_0(x)$ is the $T$-independent chemical shift and $a(x)$ is the relevant hyperfine coupling constant and, as indicated, both change only with oxygen content, $x$. For each $x$ therefore, $\chi_{\textrm{s}}$, $\chi_{\textrm{m}}$ and $^{89}$K$_{\textrm{s}}$ are linearly related to within an additive constant. We used values of $a(x)$ reported by Alloul \cite{Alloul}, while for each $x$ the additive constant was determined by matching the $^{89}\chi_{\textrm{s}}$ data to our bulk susceptibility data, $\chi_{\textrm{s}}$ \cite{Loram_IRC}. This fixed the value of the constant $\sigma_0(x)$ which differed somewhat from those of Alloul but other literature values also reflect those differences \cite{Williams2}. The overall $T$-dependence (independent of $\sigma_0(x)$) was an excellent match.

\subsection*{Acknowledgements}
We are grateful to Dr. J. R. Cooper for helpful comments on the Knight shift and spin susceptibility. We also thank Dr Martin Ryan for assistance with the x-ray diffraction analysis.

\subsection*{Author contributions}
JLT synthesized and characterized the samples, JWL carried out the specific heat measurements and the initial analysis to extract the electronic specific heat. JLT analyzed the data and wrote the paper.

\bigskip
$^\dag$ jeff.tallon@vuw.ac.nz

$^\ddag$ deceased November 2017.


\begin{references}
\small

\bibitem{Norman1} Norman, M. R., Pines, D. \& Kallin, C. The pseudogap: friend or foe of high-$T_{\textrm{c}}$. {\it Adv. Phys.} {\bf 54}, 715–733 (2005).

\bibitem{Timusk} Timusk, T. \& Statt, B. The pseudogap in high-temperature superconductors: an experimental survey. {\it Rep. Progr. Phys.} {\bf 62}, 61-122 (1999).

\bibitem{Tstar} Tallon, J. L. \& Loram,  J. W. The doping dependence of $T^*$ -- what is the real high-$T_{\textrm{c}}$ phase diagram? {\it Physica C} {\bf 349}, 53 (2001).

\bibitem{entrant} Tallon, J. L., Storey, J. G., Cooper, J. R. \& Loram, J. W. Locating the pseudogap closing point in cuprate superconductors: absence of entrant or reentrant behavior. {\it Phys. Rev. B} {\bf 101}, 174512 (2020).

\bibitem{Bourges} Fauqu\'{e}, B., Sidis, Y., Hinkov, V., Pailh\`{e}s, S. Lin, C. T., Chaud, X. \& Bourges, P. Magnetic order in the pseudogap phase of high-$T_{\textrm{c}}$ superconductors. {\it Phys. Rev. Lett.} {\bf 96}, 197001 (2006).

\bibitem{Xia} Xia, J. {\it et al.} Polar Kerr-effect measurements of the high-temperature YBa$_2$Cu$_3$O$_{6+x}$ superconductor: evidence for broken symmetry near the pseudogap temperature. {\it Phys. Rev. Lett.} {\bf 100}, 127002 (2008).

\bibitem{Hashimoto} Hashimoto, M. {\it et al.}, Particle–hole symmetry breaking in the pseudogap state of Bi2201. {\it Nature Phys.} {\bf 6}, 414-418 (2010).

\bibitem{He1} He, R.-H. {\it et al.} From a single-band metal to a high-temperature superconductor via two thermal phase transitions. {\it Science} {\bf 331}, 15791583 (2011).

\bibitem{Shekhter} Shekhter, A. {\it et al.} Bounding the pseudogap with a line of phase transitions in YBa$_2$Cu$_3$O$_{6+\delta}$.  {\it Nature} {\bf 498}, 75-77 (2013).

\bibitem{Sato} Sato, Y. {\it et al.} Thermodynamic evidence for a nematic phase transition at the onset of the pseudogap in YBa$_2$Cu$_3$O$_y$. {\it Nature Phys.} {\bf 13}, 1074-1078 (2017).

\bibitem{Loram2} Loram, J. W., Mirza, K. A., Cooper, J. R. \& Liang, W. Y. Electronic specific heat of YBa$_2$Cu$_3$O$_{6+x}$ from 1.8 to 300 K. {\it Phys. Rev. Lett.} {\bf 71}, 1740-1743 (1993).

\bibitem{Loram1}  Loram, J. W., Mirza, K. A., Cooper, J. R., Liang, W. Y. \& Wade, J. M. Electronic specific heat of YBa$_2$Cu$_3$O$_{6+x}$ from 1.8 to 300 K. {\it J. Supercon.} {\bf 7}, 243-249 (1994).

\bibitem{Loram4} Loram, J. W., Luo, J. L., Cooper, J. R., Liang, W. Y. \& Tallon, J. L. The condensation energy and pseudogap energy scale of Bi:2212 from the electronic specific heat. {\it Physica C} {\bf 341-348}, 831 (2000).

\bibitem{Williams1} Williams, G. V. M., Talion, J. L., Meinhold, R. \& J\'{a}nossy, A. $^{89}$Y NMR study of the effect of Zn substitution on the spin dynamics of YBa$_2$Cu$_4$0$_8$. {\it Phys. Rev. B} {\bf 51} 16503-16506 (1995).

\bibitem{Tallon_scattering} Tallon, J. L., Bernhard, C., Williams, G. V. M. \& Loram, J. W. Zn-induced $T_c$ Reduction in High-$T_c$ Superconductors: Scattering in the Presence
of a Pseudogap. {\it Phys. Rev. Lett.} {\bf 79}, 5294-5297 (1997).

\bibitem{Volovik} Volovik, G. E. {\it JETP Lett.} {\bf 58}, 469-473 (1993).

\bibitem{Chen} Chen, S.-D., Hashimoto, M., He, Y., Song, D., Xu, K.-J., He, J.-F., Devereaux, T. P., Eisaki, H., Lu, D.-H., Zaanen, J. \& Shen, Z.-X. Incoherent strange metal sharply bounded by a critical doping in Bi2212, {\it Science} {\bf 366}, 1099-1102 (2019).

\bibitem{Storey2} Storey, J.G. Hall effect and Fermi surface reconstruction via electron pockets in the high-$T_{\textrm{c}}$ cuprates, {\it Europhys. Lett.} {\bf 113}, 27003 (2016).

\bibitem{Kunisada} Kunisada, S. {\it et al.} Observation of small Fermi pockets protected by clean CuO$_2$ sheets of a high-$T_{\textrm{c}}$ superconductor. {\it Science} {\bf 369}, 833-838 (2020).

\bibitem{Bucher} Bucher, B., Steiner, P., Karpinski, J., Kaldis, E. \& Wachter, P. Influence of the Spin Gap on the Normal State Transport in YBa$_2$Cu$_4$0$_8$. {\it Phys. Rev. Lett.} {\bf 70}, 2012-2015 (1993).

\bibitem{Raffa} Raffa, F., Ohno, T., Mali, M., Roos, J., Brinkmann, D., Conder, K. \& Eremin, M. Isotope Dependence of the Spin Gap in YBa$_2$Cu$_4$0$_8$ as Determined by Cu NQR Relaxation. {\it Phys. Rev. Lett.} {\bf 81}, 5912-5915 (1998).

\bibitem{Vishik} Vishik, I. M. {\it et al.} Phase competition in trisected superconducting dome. {\it Proc. Nat. Acad, Sci.} {\bf 109}, 18332-18337 (2012).

\bibitem{Padamsee} Padamsee, H., Neighbor, J. E. \& Shiffman, C. A. Quasiparticle Phenomenology for Thermodynamics of Strong-Coupling Superconductors. {\it J. Low Temp. Phys.} {\bf 12}, 387-411 (1973).

\bibitem{Loram_IRC} Loram, J. W., Mirza, K. A. \& Cooper, J. R. Properties of the superconducting condensate and normal-state pseudogap in high-$T_{\textrm{c}}$ superconductors derived from the electronic specific heat. {\it IRC in Superconductivity Research Review} Cambridge University (1998).

\bibitem{Loram3} Loram, J. W., Mirza, K. A., Wade, J. M., Cooper, J. R. \& Liang, W. Y. The Electronic Specific Heat of Cuprate Superconductors. {\it Physica C} {\bf 235-240}, 134-137 (1994).

\bibitem{Tomeno} Tomeno, I., Machi, T., Tai, K., Koshizuka, N., Karnbe, S., Hayashi, A., Ueda, Y. \& Yasuoka, H. NMR study of spin dynamics at planar oxygen and copper sites in YBa$_2$Cu$_4$0$_8$. Phys. Rev. B {\bf 49}, 15327-15334 (1994).

\bibitem{Loram} Loram, J. W., Luo, J. L., Cooper, J. R., Liang, W. Y. \& Tallon, J. L. Evidence on the pseudogap and condensate from the electronic specific heat. {\it J. Phys. Chem. Solids} {\bf 62}, 59-64 (2001).

\bibitem{Junod} Wang, Y., Revaz, B., Erb, A. \& Junod, A. Direct observation and anisotropy of the contribution of gap nodes in the low-temperature specific heat of YBa$_2$Cu$_3$O$_7$. {\it Phys. Rev. B} {\bf 63}, 094508 (2001).

\bibitem{Wang} Wang, Y., Yan, J., Shan, L., Wen, H.-H., Tanabe, Y., Adachi, T. \& Koike, Y. Weak-coupling $d$-wave BCS superconductivity and unpaired electrons in overdoped La$_{2-x}$Sr$_x$CuO$_4$ single crystals. {\it Phys. Rev. B} {\bf 76}, 064512 (2007)

\bibitem{Benseman} Benseman, T. M., Cooper,  J. R. \& Balakrishnan, G. Temperature and field dependence of the intrinsic tunnelling structure in overdoped Bi$_2$Sr$_2$CaCu$_2$O$_{8+\delta}$, {\it Phys. Rev. B} {\bf 98}, 014507 (2018).

\bibitem{Sebastian} Sebastian, S. E., Harrison, N., Liang, R., Bonn, D. A., Hardy, W. N., Mielke, C. H. \& Lonzarich, G. G. Quantum Oscillations from Nodal Bilayer Magnetic Breakdown in the Underdoped High Temperature Superconductor YBa$_2$Cu$_3$O$_{6+x}$. {\it Phys. Rev. Lett.} {\bf 108}, 196403 (2012).

\bibitem{Tallon2} Tallon, J. L., Barber, F., Storey, J. G. \& Loram, J. W. Coexistence of the superconducting energy gap and pseudogap above and below the transition temperature of cuprate superconductors. {\it Phys. Rev. B} {\bf 87}, 140508(R) (2013).

\bibitem{Storey1} Storey, J. G. Incoherent superconductivity well above $T_{\textrm{c}}$ in high-$T_{\textrm{c}}$ cuprates --
harmonizing the spectroscopic and thermodynamic data. {\it New J. Phys.} {\bf 19}, 073026 (2017).

\bibitem{Kondo} Kondo, T., Malaeb, W., Ishida, Y., Sasagawa, T., Sakamoto, H., Takeuchi, T., Tohyama, T. \& Shin, S. Point nodes persisting far beyond $T_{\textrm{c}}$ in Bi2212. {\it Nature Commun.} {\bf 6}, 7699 (2015).

\bibitem{Cooper1} Cooper, J. R. \& Loram, J. W. Some Correlations Between the Thermodynamic and Transport Properties of High $T_{\textrm{c}}$ Oxides in the Normal State. {\it J. Phys. I France} {\bf 6}, 2237-2263 (1996).

\bibitem{Alloul} Alloul, H., Ohno, T. \& Mendels, P. $^{89}$Y NMR evidence for a fermi-liquid behavior in YBa$_2$Cu$_3$O$_{6+x}$. {\it Phys. Rev. Lett.} {\bf 63}, 1700-1703 (1989).

\bibitem{Williams} Williams, G. V. M., Tallon, J. L., Quilty, J. W., Trodahl, H. J. \& Flower, N. E. Absence of an Isotope Effect in the Pseudogap in YBa$_2$Cu$_4$O$_8$ as Determined by High-Resolution $^{89}$Y NMR. {\it Phys. Rev. Lett.} {\bf 80}, 377-380 (1998).

\bibitem{Crossover} Williams, G. V. M., Tallon, J. L. \& Loram, J. W. Crossover temperatures in the normal-state phase diagram of high-$T_{\textrm{c}}$ superconductors. {\it Phys. Rev. B} {\bf 58}, 15053-15061 (1998).

\bibitem{Williams2} Williams, G. V. M., Pringle, D. J. \& Tallon, J. L. Contrasting oxygen and copper isotope effects in YBa$_2$Cu$_4$O$_8$ superconducting and normal states. {\it Phys. Rev. B} {\bf 61}, 9257-9260(R) (2000).





\bibitem{Tallon7} Tallon, J. L., Bernhard, C., Binninger, U., Hofer, A., Williams, G. V. M., Ansaldo, E. J., Budnick, J. I. \& Niedermayer, Ch. In-plane anisotropy of the penetration depth due to superconductivity on the CuO chains in YBa$_2$Cu$_3$O$_{7-\delta}$, Y$_2$Ba$_4$Cu$_7$O$_{15-\delta}$ and YBa$_2$Cu$_4$O$_8$. {\it Phys. Rev. Lett.} {\bf 74}, 1008-1011 (1995).







\bibitem{Haase} Nachtigal, J., Avramovska, M., Erb, A., Pavicevic, D., Guehne, R. \& Haase, J. Temperature independent cuprate pseudogap from planar oxygen NMR. {\it Condens. Matter} {\bf 5}, 66 (2020).

\bibitem{Bernhard1} Yu, Li, Munzar, D., Boris, A. V., Yordanov, P., Chaloupka, J., Wolf, Th., Lin, C. T., Keimer, B. \& Bernhard, C. Evidence for two separate energy gaps in underdoped high-temperature cuprate superconductors from broadband infrared ellipsometry. {\it Phys. Rev. Lett.} {\bf 100}, 177004 (2008).

\bibitem{Pan} Pan, S. H., Hudson, E. W., Lang, K. M., Eisaki, H., Uchida, S. \& Davis, J. C. Imaging the effects of individual zinc impurity atoms on superconductivity in Bi$_2$Sr$_2$CaCu$_2$O$_{8+\delta}$. {\it Nature} {\bf 403}, 746-750 (2003).

\bibitem{Islam} Islam, R. S., Cooper, J. R., Loram, J. W. \& Naqib, S. H. Pseudogap and doping-dependent magnetic properties of La$_{2-x}$Sr$_x$Cu$_{1-y}$Zn$_y$O$_4$. {\it Phys. Rev. B} {\bf 81} 054511-8 (2010).

\bibitem{Obertelli} Obertelli, S. D., Cooper, J. R. \& Tallon, J. L. Systematics in the Thermoelectric power of High-$T_{\textrm{c}}$ Oxides. {\it Phys. Rev. B} {\bf 46}, 14928-14931 (1992).

\end{references}
\end{document}